\documentclass[12pt]{article}

\usepackage[]{amsmath,amscd,amsfonts,amsthm,amssymb}
\usepackage[english]{babel}

\title {Connectedness applied to closure spaces and state property systems\footnote{Published as Aerts, D., Deses, D. and Van der
Voorde, A., 2001, ``Connectedness applied to closure spaces and state property systems", {\it Journal of Electrical Engineering},
{\bf 52}, 18 - 21.}}
\author {D. Aerts, D. Deses, A. Van der Voorde}
\date {}

\newtheorem{definition}{Definition}
\newtheorem{lemma}{Lemma}
\newtheorem{proposition}{Proposition}
\newtheorem{corollary}{Corollary}

\begin{document}

\maketitle \centerline{FUND and TOPO,}
\centerline{Department of Mathematics, Brussels Free University,}
\centerline{Pleinlaan 2, B-1050 Brussels, Belgium}
\centerline{{\tt diraerts,diddesen,avdvoord@vub.ac.be}}

\begin{abstract}
\noindent In \cite{aerts} a description of a physical entity is given by
means of a state property system and in \cite{aertsann} it is
proven that any state property system is equivalent to a closure
space. In the present paper we investigate the relations between
classical properties and connectedness for closure spaces. The
main result is a decomposition theorem, which allows us
to split a state property system into a number of `pure nonclassical state property
systems' and a `totally classical state property system'.
\end{abstract}

\section{Introduction}

In \cite{aerts} a physical entity is represented by a mathematical
model called a state property system. This model contains a complete
lattice of properties of the physical entity. In \cite{aertsann}
it is shown that the lattice can viewed as the lattice of closed
sets of a closure space. We introduce the concept of classical property of the entity, and show that these correspond exactly to the
clopen (open and closed) subsets of the associated closure space. Using the concept
of connectedness for closure space we decompose the state
property system into smaller ones which will be `completely quantum mechanical'
(no classical properties) and another one which will be `totally
classical'. Finally we introduce a way to
extract the classical properties of the entity, and
interpret this in term of the closure space. Let us first introduce the basic
definitions and concepts.

\begin{definition}
\label{def:statprop} A triple $(\Sigma,{\cal L},\xi)$ is called a
state property system if $\Sigma$ is a set, ${\cal L}$ is a
complete lattice and $\xi :  \Sigma \rightarrow {\cal P}({\cal
L})$ is a function such that for $p \in \Sigma$, $0$ the minimal
element of ${\cal L}$ and $(a_i)_i \in {\cal L}$, we have:
$$0 \not\in \xi(p) \mbox{ and } \forall i \ a_i \in \xi(p)\  \Rightarrow
\wedge_i a_i \in \xi(p) \label{eq:xi_inf}$$
and for $a, b\in {\cal L}$ we have:
$  a < b \Leftrightarrow \forall r \in \Sigma:a \in \xi(r) \mbox{ then }
\ b \in \xi(r) \label{eq:xi2} $.\\
 If $(\Sigma,{\cal L},\xi)$ is a state
property system then its Cartan map is the mapping $\kappa : {\cal
L} \to {\cal P}(\Sigma)$ defined by :
\begin{equation*}
\kappa: {\cal L} \rightarrow {\cal P}(\Sigma):a \mapsto \kappa(a)
= \{p\in \Sigma \ \vert\ a \in \xi(p)\}
\end{equation*}
\end{definition}
\noindent The physical interpretation of this mathematical structure
(introduced in \cite{aerts}) is the following. Considering an
entity $S$, the set $\Sigma$ consists of states of $S$ while the
set ${\cal L}$ consists of properties of $S$. These two sets are
linked by means of a function $\xi : \Sigma \to {\cal P}({\cal
L})$ which maps a state $p$ to the set $\xi(p)$ of all properties
that are actual in state $p$. This means that the statement `a state $p$ makes
the property $a$ actual' is mathematically expressed by the formula: $a \in \xi(p)$.

\begin{definition}
\label{def:clos} A closure space $(X,{\cal F})$ consists of a set
$X$ and a family of subsets \mbox{${\cal F} \subseteq {\cal
P}(X)$} satisfying the following two conditions:
$$ \emptyset \in {\cal F} \mbox{ and } (F_i)_i \in {\cal F} \Rightarrow
\cap_iF_i \in {\cal F} $$
The closure operator corresponding to the closure space $(X,{\cal
F})$ is defined as
\begin{equation*}
cl: {\cal P}(X) \rightarrow {\cal P}(X): A \mapsto \bigcap \{F \in
{\cal F}\ \vert\ A \subseteq F\}
\end{equation*}
\end{definition}
\noindent
The following theorem shows how we can associate with each state
property system a closure space and vice versa.

\begin{proposition}
If $(\Sigma,{\cal L},\xi)$ is a state property system then
$F(\Sigma,{\cal L},\xi)=(\Sigma,\kappa(\mathcal{L}))$ is a closure space.
Conversely, if $(\Sigma,\mathcal{F})$ is a closure space then
$G(\Sigma,\mathcal{F})=(\Sigma,\mathcal{F},\xi')$, where
$\xi'(x)=\{F \in \mathcal{F}|x\in F\}$, is a state property
system.
\end{proposition}
\noindent
This is a consequence of the fact that there is an categorical
equivalence between state property systems and closure spaces, as
proven in \cite{aertsann} .

\section{Super selection rules and classical properties}

Superposition states in quantum mechanics are those
states that do not exist in classical physics and hence their
appearance is one of the important quantum characteristics. This concept
can be traced back within this general setting, by introducing the
idea of `super selection rule'. Two properties are separated by a
super selection rule iff there do not exist `superposition states'
related to these two properties.

\begin{definition}
Consider a state property system $(\Sigma,{\cal L},\xi)$. For $a,
b \in {\cal L}$ we say that $a$ and $b$ are separated by a super
selection rule, and denote $a$ ssr $b$, iff for $p \in \Sigma$ we
have:
$ a \vee b \in \xi(p) \Rightarrow a \in \xi(p) {\rm \ or}\ b \in
\xi(p)$
\end{definition}

\begin{lemma}
\label{lemma1}
Consider a state property system $(\Sigma,{\cal L},\xi)$ and its
corresponding closure system ${\cal F} = \kappa({\cal L})$. For $a
, b \in {\cal L}$ we have:
\begin{equation*}
a \ ssr\  b \Leftrightarrow \kappa(a \vee b) = \kappa(a) \cup
\kappa(b) \Leftrightarrow \kappa(a) \cup \kappa(b) \in {\cal F}
\end{equation*}
\end{lemma}
\begin{proof}
This is an easy verification.
\end{proof}
\noindent
We are ready now to introduce the concept of a `classical
property'.

\begin{definition}
Consider a state property system $(\Sigma,{\cal L},\xi)$. We say
that a property $a \in {\cal L}$ is a `classical property', if
there exists a property $a^c \in {\cal L}$ such that $a \vee a^c =
I$, $a \wedge a^c = 0$ and $a$ ssr $a^c$.
\end{definition}

\noindent Remark that for every state property system
$(\Sigma,{\cal L},\xi)$ the properties $0$ and $I$ are classical
properties. Note also that if $a \in {\cal L}$ is a classical
property, we have for $p \in \Sigma$ that $a \in \xi(p)
\Leftrightarrow a^c \notin \xi(p)$ and $a \notin \xi(p)
\Leftrightarrow a^c \in \xi(p)$. This follows immediately from the
definition of a classical property. From the previous definition and lemma \ref{lemma1} one can prove
the following lemma.

\begin{lemma}
Consider a state property system $(\Sigma,{\cal L},\xi)$. If $a
\in {\cal L}$ is a classical property, then $a^c$ is unique and is
a classical property. We call it the complement of $a$.
Further the following is satisfied:
$$(a^c)^c = a, a < b \Rightarrow b^c < a^c,\kappa(a^c) = \kappa(a)^C$$
\end{lemma}

\begin{definition}
A closure space $(X,{\cal F})$ is called connected if the only
clopen (i.e. closed and open) sets are $\emptyset$ and $X$.
\end{definition}
\noindent We can show now that the subsets that make closure systems
disconnected are exactly the subsets corresponding to classical
properties.

\begin{proposition}
\label{prop clasclop} Consider a state property system
$(\Sigma,{\cal L},\xi)$ and its corresponding closure space
$(\Sigma,\kappa({\cal L}))$. For $a \in {\cal L}$ we have:
$
a \mbox{ is classical } \Leftrightarrow \kappa(a)$ is
clopen.
\end{proposition}
\begin{proof}
From the previous lemmas it follows that if $a$ is
classical, then $\kappa(a)$ is clopen. So now consider a clopen
subset $\kappa(a)$ of $\Sigma$. This means that $\kappa(a)^C$ is
closed, and hence that there exists a property $b \in {\cal L}$
such that $\kappa(b) = \kappa(a)^C$. We clearly have $a \wedge b =
0$ since there exists no state $p \in \Sigma$ such that $p \in
\kappa(a)$ and $p \in \kappa(b)$. Since $\Sigma = \kappa(a) \cup
\kappa(b)$ we have $a \vee b = I$. Further we have that for an
arbitrary state $p \in \Sigma$ we have $a \in \xi(p)$ or $b \in
\xi(p)$ which shows that $a$ ssr $b$. This proves that $b = a^c$
and that $a$ is classical.
\end{proof}

\noindent This means that the classical properties correspond
exactly to the clopen subsets of the closure system.

\begin{corollary}
\label{cor quantum} Let $(\Sigma,{\cal L},\xi)$ be a state propery system. T.F.A.E.
\\ (1) The properties $0$ and $I$ are the only classical ones. \\
(2) $F(\Sigma,{\cal L},\xi) = (\Sigma, \kappa({\cal L}))$ is a connected closure
space.
\end{corollary}

\begin{definition}
A state property system $(\Sigma,{\cal L},\xi)$ is called a `pure nonclassical state property
system' if the properties $0$ and $I$ are the only classical
properties.
\end{definition}

\begin{proposition}
Let $(\Sigma,{\cal F})$ be a closure space. T.F.A.E. \\ (1)
$(\Sigma,\cal F)$ is a connected closure space. \\ (2)
$G(\Sigma,{\cal F}) = (\Sigma,{\cal F},\xi)$ is a pure nonclassical state
property system.
\end{proposition}
\begin{proof}
Let $(\Sigma,{\cal F})$ be a connected state property system. Then
$\emptyset$ and $\Sigma$ are the only clopen sets in
$(\Sigma,{\cal F})$. Since the Cartan map associated to $\xi$ is
given by $\kappa : {\cal F} \to {\cal P}(\Sigma) : F \mapsto F$,
we have $\kappa(\emptyset) = \emptyset$ and $\kappa(\Sigma) =
\Sigma$. Applying proposition~\ref{prop clasclop}, we find that
$\emptyset$ and $\Sigma$ are the only classical properties of
${\cal F}$. Conversely, let $G(\Sigma,{\cal F}) = (\Sigma,{\cal
F},\xi)$ be a pure nonclassical state property system. Then by
corollary~\ref{cor quantum}, $(\Sigma,{\cal F}) = FG(\Sigma,{\cal
F})$ is a connected closure space.
\end{proof}

\section{Decomposition theorem}

As for topological spaces, every closure space can be decomposed
uniquely into connected components. In the following we say that,
for a closure space $(X,{\cal F})$, a subset $A \subseteq X$ is
connected if the induced subspace is connected. It can be shown
that the union of any family of connected subsets having at least
one point in common is also connected. So the component of an
element $x \in X$ defined by
$ K_{\bf Cls}(x)=\bigcup \{A\subseteq X \ | \ x \in A, A \hbox{
connected }\}$
is connected and therefore called the connection component of $x$.
Moreover, it is a maximal connected set in $X$ in the
sense that there is no connected subset of $X$ which properly
contains $K_{\bf Cls}(x)$. From this it follows that for closure
spaces $(X,{\cal F})$ the set of all distinct connection
components in $X$ form a partition of $X$.
In the following we
will decompose state property systems similarly into
different components.

\begin{proposition}
Let $(\Sigma,{\cal L},\xi)$ be a state property system and let
$(\Sigma,\kappa({\cal L}))$ be the corresponding closure space.
Consider the equivalence relation on $\Sigma$ given by:
$ p \sim q \Leftrightarrow K_{\bf Cls}(p)=K_{\bf Cls}(q) $
with equivalence classes $\Omega=\{\omega(p)|p\in \Sigma\}$. If
$\omega \in \Omega$ we define:
\begin{eqnarray*}
\Sigma_\omega &=&\omega  = \{p \in \Sigma \ | \ \omega(p) = \omega\} \\
s(\omega)&=&s(\omega(p))=a, \textrm{ such
that } \kappa(a)=\omega(p) \\ {\cal
L}_\omega&=&[0,s(\omega)]=\{a\in {\cal L} \ | \ 0 \leq a \leq s(\omega)
\}\subset {\cal L}\\ \xi_\omega &:&\Sigma_\omega\to {\cal P}({\cal
L}_\omega):p \mapsto \xi(p)\cap {\cal L}_\omega
\end{eqnarray*}
then $(\Sigma_\omega,{\cal L}_\omega,\xi_\omega)$ is a state property
system.
\end{proposition}

\begin{proof}
Since ${\cal L}_\omega$ is a sublattice (segment) of ${\cal L}$,
it is a complete lattice with maximal element $I_\omega=s(\omega)$
and minimal element $0_\omega=0$. Let $p \in \Sigma_\omega$. Then
$0 \not \in \xi(p)$. So $0 \not \in \xi(p)\cap {\cal
L}_\omega=\xi_\omega(p)$. If $a_i \in \xi_\omega(p), \ \forall i$,
then $a_i \in {\cal L}_\omega$ and $a_i\in \xi(p), \ \forall i$.
Hence $\wedge a_i \in {\cal L}_\omega \cap \xi(p) =
\xi_\omega(p)$. Finally, let $a,b \in {\cal L}_\omega$ with
$a<_\omega b$ and let $r\in \Sigma_\omega$. If $a \in
\xi_\omega(r)$, then $a\in {\cal L}_\omega$ and $a\in \xi(r)$,
thus $b\in {\cal L}_\omega$ and $b\in \xi(r)$. So $b\in
\xi_\omega(r)$. Conversely, if $a,b \in {\cal L}_\omega$ and
$\forall r\in \Sigma_\omega: a\in \xi_\omega(r) \Rightarrow b\in
\xi_\omega(r)$ then we consider a $q$ such that $a\in \xi(q)$ ($q$
must be in $\Sigma_\omega$ by definition of ${\cal L}_\omega$).
Then $a\in \xi_\omega(q)$ implies that $b\in \xi_\omega(q)$. So
$b\in \xi(q)$ and $a < b$. Thus $a<_\omega b$.
\end{proof}

\noindent Moreover we can show that the above introduced state property
systems $(\Sigma_{\omega},{\cal L}_{\omega},\xi_{\omega})$  have
no proper classical properties, and hence are pure nonclassical state
property systems.

\begin{proposition}
Let $(\Sigma,{\cal L},\xi)$ be a state property system. If $\omega \in \Omega$, then $(\Sigma_\omega,{\cal
L}_\omega,\xi_\omega)$ is a pure nonclassical state property system.
\end{proposition}

\begin{proof}
If $a$ is classical element of ${\cal L}_\omega$, then $\kappa(a)$
must be a clopen set of the associated closure space
$(\Sigma_{\omega},\kappa({\cal L}_\omega))$ which
is a connected subspace of $(\Sigma,\kappa({\cal L}))$. Hence
there are no proper classical elements of ${\cal L}_\omega$.
\end{proof}

\begin{proposition}Let $(\Sigma,{\cal L},\xi)$ be a state property system. If we introduce the following :
\begin{eqnarray*}
\Omega&=&\{\omega(p) \ | \ p\in \Sigma\}\\ \mathcal{C}&=&\{\vee
s(\omega_i) \ | \ \omega_i\in \Omega\} \\ \eta&:&\Omega \to
{\cal P}(\mathcal{C}):\omega=\omega(p)\mapsto \xi(p)\cap \mathcal{C}
\end{eqnarray*}
then $(\Omega,{\cal C},\xi)$ is a state property system.
\end{proposition}

\begin{proof}
First we remark that $\eta$ is well defined because if
$\omega(p)=\omega(q)$, then $\xi(p) \cap {\cal C} = \xi(q) \cap
{\cal C}$. Indeed, if $\vee s(\omega_i)\in \xi(p)$ then $p \in
\kappa(\vee s(\omega_i)) = cl( \cup\omega_i)$ in the corresponding
closure space $(\Sigma,\kappa({\cal L}))$. Since
$cl(\cup\omega_i)$ is not connected we have that $K_{\bf
Cls}(p)=\omega(p)=\omega(q)\subset cl(\cup\omega_i)$ so $q \in cl(
\cup\omega_i) = \kappa(\vee s(\omega_i))$ and $\vee s(\omega_i)\in
\xi(q)$. Now, since $\mathcal{C}$ is a sublattice of ${\cal L}$ it
is a complete lattice with $1_\mathcal{C}=1$ and
$0_\mathcal{C}=0$. By definition $\mathcal{C}$ is generated by its
atoms $\{s(\omega) \ | \ \omega\in \Omega\}$. Clearly $0\not \in
\eta(\omega(p))$ because $0 \not \in \xi(p)$. If $a_i\in
\eta(\omega(p))=\xi(p) \cap \mathcal{C}, \ \forall i$, then
$\wedge a_i\in \xi(p)\cap \mathcal{C}=\eta(\omega(p))$. Finally,
let $a,b \in {\cal C}$  with $a<_\mathcal{C} b$. Let $\omega(p)
\in \Omega$ with $a \in \eta(\omega(p))$. Thus  $a \in \xi(p)$. $a
<_{\cal C} b$ implies $a < b$. So we have $b \in \xi(p) \cap {\cal
C} = \eta(\omega(p))$. Conversely, let $a,b\in \mathcal{C}$ and
assume that $\forall p\in \Sigma : a\in\eta(\omega(p))\Rightarrow
b \in\eta(\omega(p))$. Then we have $\forall p\in \Sigma:
a\in\xi(p)\Rightarrow b \in\xi(p)$. Thus $a< b$ and
$a<_\mathcal{C}b$.
\end{proof}

\begin{proposition}
$(\Omega,\mathcal{C},\eta)$ is a totally classical state property
system, in the sense that the only quantum segments (i.e. segments
with no proper classical elements) are trivial, i.e.
$\{0,s(\omega)\}$.
\end{proposition}

\begin{proof}
Suppose $[0,a]$ is a quantumsegment of $\mathcal{C}$, then in the
corresponding closure space $(\Sigma,\kappa({\mathcal L}))$ the
subset $\kappa(a)$ is connected hence $\kappa(a)\subset\omega$ for
some $\omega\in \Omega$, hence $a<s(\omega)$. Since $s(\omega)$ is an atom, $a = s(\omega)$.
Thus $[0,a] = \{0,s(\omega)\}$.
\end{proof}

\begin{corollary}
The closure space associated with $(\Omega,\mathcal{C},\eta)$ is a
totally disconnected closure space.
\end{corollary}
\noindent
Summarizing the previous results we get:

\begin{proposition}
Any state property system $(\Sigma,{\cal L},\xi)$ can be decomposed into: a number
of pure nonclassical state property systems $(\Sigma_\omega,{\cal
L}_\omega,\xi_\omega),\omega \in \Omega$ and a totally classical
state property system $(\Omega,\mathcal{C},\eta)$
\end{proposition}

\section{The classical part of a state property system}

In this section we want to show how it is possible to extract the
classical part of a state property system. First of all we have to
define the classical property lattice related to the entity $S$
that is described by the state property system $(\Sigma,{\cal
L},\xi)$.

\begin{definition} [Classical property lattice]
Consider a state property system $(\Sigma,{\cal L},\xi)$. We call
${\cal C}' = \{\wedge_ia_i \vert a_i \ {\rm is \ a \ classical \
property }\}$ the classical property lattice corresponding to the
state property system $(\Sigma,{\cal L},\xi)$.
\end{definition}

\begin{proposition}${\cal C}'$ is a complete lattice with the partial order relation
and infimum inherited from ${\cal L}$ and the supremum defined as
follows: for $a_i \in {\cal C}'$,  $\vee_ia_i = \wedge_{b \in
{\cal C}', a_i \le b\ \forall i}\ b$.
\end{proposition}

\noindent Remark that the supremum in the lattice ${\cal C}'$ is
not the one inherited from ${\cal L}$.

\begin{proposition}
Consider a state property system $(\Sigma,{\cal L},\xi)$. Let $\xi'(q)=\xi(q)\cap
{\cal C}'$ for $q\in \Sigma$, then $(\Sigma,\mathcal{C}',\xi')$ is
a state property system which we shall refer to as the classical
part of $(\Sigma,{\cal L},\xi)$.
\end{proposition}

\begin{proof}
Clearly $0 \not\in \xi'(p)$ for $p\in \Sigma$. Consider $a_i \in
\xi'(p)\ \forall i$. Then $a_i \in \xi(p) \cap {\cal C}' \ \forall
i$, from which follows that $\wedge_ia_i \in \xi(p) \cap {\cal
C}'$ and hence $\wedge_ia_i \in \xi'(p)$. Consider $a, b \in {\cal
C}'$. Let us suppose that $a \leq b$ and consider $r \in \Sigma$
such that $a \in \xi'(r)$. This means that $a \in \xi(r) \cap
{\cal C}'$. From this follows that $b \in \xi(r) \cap {\cal C}'$
and hence $b \in \xi'(r)$. On the other hand let us suppose that
$\forall r \in \Sigma:a \in \xi'(r) {\rm \ then}\ b \in \xi'(r)$.
Since $a, b \in {\cal C}'$, this also means that $\forall r \in
\Sigma:a \in \xi(r) {\rm \ then}\ b \in \xi(r)$. From this follows
that $a \le b$.
\end{proof}

\noindent
Since $(\Sigma,\mathcal{C}',\xi')$ is a state property system, it
has a corresponding closure space $(\Sigma,\kappa(\mathcal{C}'))$.
In order to check some property of this space we introduce the
following concepts.

\begin{definition}
Let $(X,\mathcal{F})$ be a closure space and $\mathcal{B}\subset
\mathcal{F}$. $\mathcal{B}$ is called a base of $(X,\mathcal{F})$
iff $\forall F\in \mathcal{F}:\exists B_i\in \mathcal{B}:F=\cap
B_i$. $(X,\mathcal{F})$ is called weakly zero-dimensional iff
there is a base consisting of clopen sets.
\end{definition}

\begin{proposition}
The closure space $(\Sigma,\kappa(\mathcal{C}'))$ corresponding to
the state property system $(\Sigma,\mathcal{C}',\xi')$  is weakly
zero-dimensional.
\end{proposition}

\begin{proof}
To see this recall that $a$ is classical iff $\kappa(a)$ is clopen
in $(\Sigma,\kappa(\mathcal{L}))$, hence $\kappa(\mathcal{C}')$ is
a family of closed sets on $\Sigma$ which consists of all
intersections of the clopen sets of
$(\Sigma,\kappa(\mathcal{L}))$.
\end{proof}
\noindent
In general the lattice of $(\Sigma,\mathcal{C}',\xi')$ does not
need to be atomistic, hence it is different from the totally
classical state property system $(\Omega,\mathcal{C},\eta)$
associated with $(\Sigma,{\cal L},\xi)$.

\noindent
{\bf Didier Deses} has presented this subject at SCAM 2001. He is
a research assistant of the fund for scientific research flanders
and Phd. student at the free university of Brussels, his
supervisor is Prof. Eva Colebunders. His main interest lies in
General and Categorical Topology.
\end{document}